\documentclass[aps,prb,twocolumn,superscriptaddress,eqsecnum,showpacs,showkeys,amsmath,amssymb]{revtex4}

\usepackage{amsfonts}
\usepackage{amssymb,amsmath}
\usepackage{graphicx}
\usepackage{dcolumn}
\usepackage{bm}
\usepackage{color}

\begin{document}

\title{Magnetism-driven ferroelectricity in spin-1/2 $XY$ chains}

\author{Oleg Menchyshyn}
\affiliation{Institute for Condensed Matter Physics,
          National Academy of Sciences of Ukraine,
          Svientsitskii Street 1, 79011 L'viv, Ukraine}

\author{Vadim Ohanyan}
\affiliation{Department of Theoretical Physics,
          Yerevan State University,
          Alex Manoogian 1, 0025 Yerevan, Armenia}
\affiliation{Abdus Salam International Centre for Theoretical Physics,
          Strada Costiera 11, 34151 Trieste, Italy}

\author{Taras Verkholyak}
\affiliation{Institute for Condensed Matter Physics,
          National Academy of Sciences of Ukraine,
          Svientsitskii Street 1, 79011 L'viv, Ukraine}

\author{Taras Krokhmalskii}
\affiliation{Institute for Condensed Matter Physics,
          National Academy of Sciences of Ukraine,
          Svientsitskii Street 1, 79011 L'viv, Ukraine}

\author{Oleg Derzhko}
\affiliation{Institute for Condensed Matter Physics,
          National Academy of Sciences of Ukraine,
          Svientsitskii Street 1, 79011 L'viv, Ukraine}
\affiliation{Department for Metal Physics,
          Ivan Franko National University of L'viv,
          Kyryla \& Mephodiya Street 8, 79005 L'viv, Ukraine}
\affiliation{Abdus Salam International Centre for Theoretical Physics,
          Strada Costiera 11, 34151 Trieste, Italy}

\date{\today}



\begin{abstract}
We illustrate the magnetoelectric effect conditioned by the Katsura-Nagaosa-Balatsky (KNB) mechanism
within the frames of exactly solvable spin-1/2 $XY$ chains.
Due to three-spin interactions which are present in our consideration,
the magnetization (polarization) is influenced by the electric (magnetic) field 
even in the absence of the magnetic (electric) field.
We also discuss a magnetoelectrocaloric effect
examining the entropy changes under the isothermal varying of the magnetic or/and electric field.
\end{abstract}

\maketitle

\section{Introduction}
\label{sec1}

Magnetically induced ferroelectricity is an important and highly interesting phenomena in condensed matter physics
which attracts essential amount of attention both form experimentalists and theoreticians.\cite{mee1,mee2}
In general, 
the phenomenon is a part of the so-called magnetoelectric effect, 
the intercoupling of magnetization and polarization in matter, 
which in its most prominent form can be briefly described 
as the magnetization dependence on the electric field and the polarization dependence on the magnetic field. 
The importance of the topic is not only limited to the highly nontrivial and rich physics, 
but it has very promising practical applications in future electronic devices.\cite{app1,app2}

Among the various physical realizations of the magnetoelectric effect, 
the ferroelectricity of spin origin is particularly important for spin-related electronics.
There are several mechanisms of the coupling 
between local spins (magnetic moments) of the magnetic materials and the local polarization of the lattice cell.\cite{mee1,mee2,app1,app2} 
The one, which we are going to exploit in the present study is called the Katsura-Nagaosa-Balatsky (KNB) mechanism,\cite{KNB1,KNB2} 
which is based on a so-called inverse Dzyaloshinskii-Moriya model or spin current model. 
This mechanism gives for the polarization (dipole moment) of the bond 
$\mathbf{p}_{ij}$ 
connecting two spins ${\mathbf{s}}_{i}$ and ${\mathbf{s}}_{j}$
the following expression:
\begin{eqnarray}
\label{101}
\mathbf{p}_{ij}\propto \left[\mathbf{e}_{ij}\times \left[{\bf{s}}_{i}\times {\bf{s}}_{j}\right]\right],
\end{eqnarray}
where $\mathbf{e}_{ij}$ is the unit vector pointing form $i$-th site to $j$-th site. 
In terms of the spin current flowing between the magnetic sites
\begin{eqnarray}
\label{102}
\mathbf{j}_{ij}\propto [\mathbf{s}_{i}\times \mathbf{s}_{j}],
\end{eqnarray}
the bond polarization in Eq.~(\ref{101}) can be rewritten as
\begin{eqnarray}
\label{103}
\mathbf{p}_{ij}\propto [\mathbf{e}_{ij}\times \mathbf{j}_{ij}].
\end{eqnarray}
According to Eqs.~(\ref{101}) -- (\ref{103}),
a spiral spin order may generate a macroscopic polarization of electronic (spin) origin.

At the present time a number of compounds is known 
which give evidence of magnetically-driven ferroelectricity and magnetoelectric effect 
due to KNB mechanism.\cite{mee2,LiCu2O21,LiCu2O22,LiCu2O23,LiCu2O24,LiCuVO41,LiCuVO42,LiCuVO43,CuCl2}
The simplest quantum spin model exhibiting the spiral magnetic order is a spin-1/2 $J_1$-$J_2$ chain 
with ferromagnetic nearest-neighbor coupling ($J_1$) and antiferromagnetic next-nearest-neighbor coupling ($J_2$) 
which is considered to be more or less realistic model for such materials as 
LiCu$_2$O$_2$,\cite{LiCu2O21,LiCu2O22,LiCu2O23,LiCu2O24} 
LiCuVO$_4$,\cite{LiCuVO41,LiCuVO42,LiCuVO43} 
and 
CuCl$_2$\cite{CuCl2} 
to mention just few of them. 

In the present paper we are going to examine an exactly solvable model exhibiting a nontrivial magnetoelectric effect. 
The introduced quantum spin-chain model contains, in addition to the common two-spin interactions, 
the three-spin interactions
and the bond polarization is of the spin origin according to KNB scenario (\ref{101}) -- (\ref{103}).
The spin system under consideration is a variant of the famous spin-1/2 $XY$ chain\cite{lsm}  
(Suzuki models\cite{suzuki1,suzuki2})
which is solvable by the Jordan-Wigner fermionization.
Such free-fermion spin models are quite popular 
as they provide suitable playgrounds for the exact description of various phenomena 
in strongly correlated systems.\cite{rossler,drd,titvinidze,lou,zvyagin,kdsv08,kdsv09,drd2,lima,topilko}
It is also in order to mention here similar recent studies 
on the exact treatment of the magnetism-driven ferroelectricity in quantum spin chains
which, however, deal with the two-spin interactions only in the $XXZ$ Heisenberg\cite{vadim,thakur} or compass models.\cite{oles}
The polarization for those models is influenced by the magnetic field only in the presence of the electric field,
but the polarization is zero at zero electric field independently on the magnetic field.
In other words, those models exhibit only the so-called {\it trivial} magnetoelectric effect.
Contrarily to those previously studied models,
the ones considered below show the {\it nontrivial} magnetoelectric effect:
The polarization is affected by the magnetic field even at zero electric field.

The rest of the paper is organized as follows.
In Sec.~\ref{sec2} we discuss the spin-chain models of magnetism-driven ferroelectricity which can be treated exactly
by the Jordan-Wigner transformation to free fermions.
Exact solutions are presented in Sec.~\ref{sec3}.
The essential feature of the models at hand is the three-spin interactions.
For the sake of simplicity we distinguish two types of the three-spin interactions,
the $XZY-YZX$ ones and the $XZX+YZY$ ones,
which are considered separately in Sec.~\ref{sec4} and Sec.~\ref{sec5}, respectively.
Our main focus is on the ground-state magnetization and polarization (zero-temperature properties)
as well as on the magnetoelectrocaloric effect (finite-temperature properties).
We summarize our findings and sketch perspectives for further work in Sec.~\ref{sec6}.
Details of some calculations are presented in two appendices.

\section{Spin-1/2 $XY$ chains augmented by KNB mechanism}
\label{sec2}

In this paper, 
we consider $N$ spins 1/2 placed on a one-dimensional linear-chain lattice.
We start with the ``bare'' Hamiltonian of the spin system
\begin{eqnarray}
\label{201}
H_0=H_J+H_E+H_K+H_{\rm{Z}},
\end{eqnarray}
which contains, in general, the ordinary two-spin isotropic $XY$ interactions,
\begin{eqnarray}
\label{202}
H_J=J\sum_n\left(s_n^xs_{n+1}^x + s_n^ys_{n+1}^y\right)
\end{eqnarray}
(we may set $J>0$ without sacrificing generality),
the three-spin interactions of two types,\cite{rossler,drd}
\begin{eqnarray}
\label{203}
H_E=E\sum_n\left(s_n^xs_{n+1}^zs_{n+2}^y - s_n^ys_{n+1}^zs_{n+2}^x\right),
\nonumber\\
H_K=K\sum_n\left(s_n^xs_{n+1}^zs_{n+2}^x + s_n^ys_{n+1}^zs_{n+2}^y\right),
\end{eqnarray}
as well as the Zeeman interaction of the spins with a $z$-aligned external magnetic field ${\cal{H}}$,
\begin{eqnarray}
\label{204}
H_{\rm{Z}}= - {\cal{H}}\sum_n s_n^z.
\end{eqnarray}
Boundary conditions are not considered explicitly for thermodynamically large systems $N\to\infty$.

In our study,
we assume that the spin model arises from a more fundamental electronic model
for which the KNB scenario\cite{KNB1} holds.
This means
that the electric polarization of the bond between the neighboring sites $n$ and $n+1$
to be denoted by
${\bf p}_{n,n+1}$
is determined by the spin current ${\bf j}_{n,n+1}$ according to Eqs.~(\ref{101}) -- (\ref{103}).
In our model, the chain runs along the $x$-direction in the real space,
i.e., $\mathbf{e}_{ij}=(1,0,0)$ in Eqs.~(\ref{101}), (\ref{103}).
Furthermore, the $x$-, $y$-, and $z$-axis in the real space and in the spin space coincide.
Then
${\bf p}_{n,n+1}\propto (0,-j^z_{n,n+1},j^y_{n,n+1})$,
i.e., the bond polarization has zero $x$-component.

The spin-current operator satisfies the lattice version of the continuity equation,
${\rm{d}}s^{\alpha}_{n}/{\rm{d}}t=-{\rm{div}}j^{\alpha}_{n}$, $\alpha=x,y,z$,
where the discrete divergence operator acts in the real space.
Consider at first the isotropic $XY$ interactions only
[i.e., $E=K=0$ in Eqs.~(\ref{201}) -- (\ref{204})].
For the $z$-component of the current through the bond we find:
\begin{eqnarray}
\label{205}
\frac{{\rm{d}} s^z_{n}}{{\rm{d}}t}=-{\rm{i}}\left[s_n^z,H_J+H_{\rm{Z}}\right]
=-\left(j^z_{n+\frac{1}{2}}-j^z_{n-\frac{1}{2}}\right)
=-{\rm{div}}j^{z}_{n},
\nonumber\\
j_{n+\frac{1}{2}}^z\equiv J\left(s^x_{n}s^y_{n+1}-s^y_{n}s^x_{n+1}\right).
\;\;\;
\end{eqnarray}
If the spin-1/2 isotropic $XY$ chain comes from an underlying electronic model for which KNB mechanism works,
then each bond has the polarization ${\bf p}_{n,n+1}\propto (0,-j^z_{n,n+1},j^y_{n,n+1})$
with $j^z_{n,n+1}=j_{n+\frac{1}{2}}^z$ given in Eq.~(\ref{205})
which may manifest itself in the presence of electric field.
Let ${\cal{E}}$ be a $y$-aligned external uniform electric field
[i.e., the electric field vector is $(0,{\cal{E}},0)$].
Then the Hamiltonian of the model $H_0$ has to be supplemented with the term 
$-{\cal{E}}\sum_n p^y_{n,n+1}\propto {\cal{E}}\sum_n j^z_{n,n+1}$
and we arrive at\cite{vadim} 
\begin{eqnarray}
\label{206}
H=J\sum_n\left[s_n^xs_{n+1}^x + s_n^ys_{n+1}^y
+{\cal{E}}\left(s_n^xs_{n+1}^y - s_n^ys_{n+1}^x\right)\right]
\nonumber\\
-{\cal{H}}\sum_n s_n^z.
\;\;\;
\end{eqnarray}
One may calculate 
the magnetic moment $M^z=\sum_n\langle s_n^z\rangle$
and 
the polarization $P^y\propto -J\sum_n\langle s_n^xs_{n+1}^y-s_n^ys_{n+1}^x\rangle$ 
for spin model (\ref{206})
using the standard Jordan-Wigner fermionization method,\cite{lsm}
see Ref.~\onlinecite{vadim} and Sec.~\ref{sec3}.

Next we consider the model with the three-spin interactions, see Eq.~(\ref{203}).
Placing the system in a uniform external electric field $(0,{\cal{E}},0)$ 
we have to add to the Hamiltonian $H_0$ the term $-{\cal{E}}P^y$.
Following previous studies on the magnetoelectric effect in the spin-1/2 $J_1$-$J_2$ chain,\cite{J1J21,J1J22,J1J23}
we take for the required polarization operator the following expression:
$P^y=\sum_n p^y_{n,n+1}\propto -\sum_n j^z_{n,n+1}$
with $j^z_{n,n+1}=j_{n+\frac{1}{2}}^z$ given in Eq.~(\ref{205}).
Then the Hamiltonian of the spin system becomes
\begin{eqnarray}
\label{207}
H=J\sum_n\left[s_n^xs_{n+1}^x + s_n^ys_{n+1}^y
+{\cal{E}}\left(s_n^xs_{n+1}^y - s_n^ys_{n+1}^x\right)\right]
\nonumber\\
+E\sum_n\left(s_n^xs_{n+1}^zs_{n+2}^y - s_n^ys_{n+1}^zs_{n+2}^x\right)
\nonumber\\
+K\sum_n\left(s_n^xs_{n+1}^zs_{n+2}^x + s_n^ys_{n+1}^zs_{n+2}^y\right)
\nonumber\\
-{\cal{H}}\sum_n s_n^z.
\;\;\;
\end{eqnarray}
As previously,
the magnetic moment $M^z=\sum_n\langle s_n^z\rangle$
and 
the polarization $P^y\propto -J\sum_n\langle s_n^xs_{n+1}^y-s_n^ys_{n+1}^x\rangle$ 
for spin model (\ref{207})
can be calculated rigorously within the frames of the Jordan-Wigner fermionization approach,\cite{lsm}
see Sec.~\ref{sec3}.
It should be also noted here,
that models similar to the one given in Eq.~(\ref{207}) have been considered recently
[compare Eq.~(\ref{207})
to Eq.~(2.1) of Ref.~\onlinecite{vadim}
or 
to Eq.~(2.2) of Ref.~\onlinecite{oles}],
however, they contain the two-spin interactions only.
As it is shown below,
the three-spin interactions lead to new features of the magnetoelectric effect.

Interestingly,
for the spin model defined in Eqs.~(\ref{201}) -- (\ref{204})
we may also consider the three-spin-interaction contribution to the spin current
in KNB formula for the bond polarization (\ref{103}) 
(see Ref.~\onlinecite{sirker})
and remain to face an exactly solvable spin model.
Indeed,
using the continuity equation,
for the required $z$-component of the spin-current operator we find:
\begin{eqnarray}
\label{208}
\frac{{\rm{d}} s^z_{n}}{{\rm{d}}t}=-{\rm{i}}\left[s_n^z,H_E+H_K\right]
=-\frac{{\sf{j}}^z_{n+1}-{\sf{j}}^z_{n-1}}{2}
=-{\rm{div}}{\sf{j}}^{z}_{n},
\nonumber\\
{\sf{j}}_{n+1}^z
\equiv
-2E\left(s^x_{n}s^z_{n+1}s^x_{n+2}+s^y_{n}s^z_{n+1}s^y_{n+2}\right)
\nonumber\\
+2K\left(s^x_{n}s^z_{n+1}s^y_{n+2}-s^y_{n}s^z_{n+1}s^x_{n+2}\right).
\;\;\;
\end{eqnarray}
Therefore in the presence of the electric field $(0,{\cal{E}},0)$ 
we face the Hamiltonian similar to the one given in Eq.~(\ref{207})
but with
\begin{eqnarray}
\label{209}
E \to E+2K{\cal{E}},
\;\;\;
K \to K-2E{\cal{E}}.
\end{eqnarray}
Clearly, all averages are calculated now with this new Hamiltonian (\ref{207}), (\ref{209}).
The magnetic moment is given by $M^z=\sum_n\langle s_n^z\rangle$,
whereas the polarization now is the sum of the two terms,
$P^y\propto -J\sum_n\langle s_n^xs_{n+1}^y-s_n^ys_{n+1}^x\rangle $ 
and
${\sf{P}}^y\propto \sum_n
(2E\langle s^x_{n}s^z_{n+1}s^x_{n+2}+ s^y_{n}s^z_{n+1}s^y_{n+2}\rangle
-2K\langle s^x_{n}s^z_{n+1}s^y_{n+2}-s^y_{n}s^z_{n+1}s^x_{n+2}\rangle)$,
see Eq.~(\ref{208}).
Again these quantities can be calculated rigorously 
using the Jordan-Wigner transformation to fermions,\cite{lsm}
see Sec.~\ref{sec3}.

In summary,
basing on the KNB scenario for magnetism-driven ferroelectricity (\ref{101}) -- (\ref{103})
and choosing specific exchange interactions, lattice geometry, and direction of external fields
we arrive at simple spin models which are expected to mimic basic features of some multiferroics.
In contrast to the considered previously similar exactly-solvable models of magnetism-driven ferroelectricity,\cite{vadim,thakur,oles}
we have included into consideration also the three-spin interactions.
More realistic description of multiferroics of spin origin leads to more sophisticated spin models 
(see, e.g., Refs.~\onlinecite{J1J21,J1J22,J1J23,sirker}) 
and their further analysis involves approximations.
The merit of the introduced models is their exact solvability.
In the following sections we examine rigorously some aspects of the magnetoelectric effect 
in spin models defined in Eq.~(\ref{206}), Eq.~(\ref{207}), and Eqs.~(\ref{207}), (\ref{209}).

To close the section,
we make a remark concerning the three-spin interactions which are present in Eq.~(\ref{207}).
Obviously, 
the spin model with two-spin interactions only (\ref{206}) describes the influence 
of ${\cal{E}}$ on $M^z$
or
of ${\cal{H}}$ on $P^y$.
However,
from symmetry arguments it is clear
that 
$M^z=0$ if ${\cal{H}}=0$ independently on ${\cal{E}}$
or
$P^y=0$ if ${\cal{E}}=0$ independently on ${\cal{H}}$,
see the paragraph after Eq.~(\ref{306}) in Sec.~\ref{sec3}.
In contrast,
in the presence of the three-site interactions (\ref{203})
these symmetry arguments do not work any more
and one may expect a nontrivial magnetoelectric effect 
when 
$M^z$ is influenced by ${\cal{E}}$ even for ${\cal{H}}=0$
or 
$P^y$ is influenced by ${\cal{H}}$ even for ${\cal{E}}=0$.
Rigorous calculations reported in Secs.~\ref{sec4} and \ref{sec5} support these expectations.

\section{Exact solutions}
\label{sec3}

Spin-1/2 models given in Eqs.~(\ref{206}), (\ref{207}), and (\ref{207}), (\ref{209}) 
are exactly solvable via the Jordan-Wigner transformation to fermions,\cite{lsm}
see Appendix~\ref{a}.
In fermionic picture we face noninteracting spinless fermions with known energies $\epsilon_\kappa$
and thus many statistical mechanical calculations can be easily carried out.
For the Helmholtz free energy per site we find\cite{kdsv08,topilko}
\begin{eqnarray}
\label{301}
f(T,{\cal{H}},{\cal{E}})
=
-\frac{T}{2\pi}\int_{-\pi}^{\pi}{\rm{d}}\kappa
\ln\left(2\cosh\frac{\epsilon_\kappa}{2T}\right),
\end{eqnarray}
where for model (\ref{207})
\begin{eqnarray}
\label{302}
\epsilon_\kappa
=-{\cal{H}}
+J \cos\kappa + J{\cal{E}} \sin\kappa
\nonumber\\
-\frac{E}{2}\sin(2\kappa)-\frac{K}{2}\cos(2\kappa).
\end{eqnarray}
In the case of model (\ref{206}) one has to put $E=K=0$ in Eq.~(\ref{302}),
whereas in the case of model (\ref{207}), (\ref{209})
one has to make the replacement (\ref{209}) in Eq.~(\ref{302}).

In what follows we are interested in the magnetization and the polarization for the spin models at hand.
For the magnetization per site we have
\begin{eqnarray}
\label{303}
m=m^z=\frac{1}{N}\sum_n\langle s_n^z\rangle
=\frac{1}{2\pi}\int_{-\pi}^{\pi}{\rm{d}}\kappa n_\kappa -\frac{1}{2},
\nonumber\\
n_\kappa=\frac{1}{1+e^{\frac{\epsilon_\kappa}{T}}}, 
\;\;\;
0\le n_\kappa\le 1.
\end{eqnarray}
For the polarization per site for spin model (\ref{207}) we have
\begin{eqnarray}
\label{304}
p=p^y\propto-\frac{J}{N}\sum_n\langle s_n^xs_{n+1}^y-s_n^ys_{n+1}^x\rangle
\nonumber\\
=-\frac{J}{2\pi}\int_{-\pi}^{\pi}{\rm{d}}\kappa \sin\kappa\, n_\kappa.
\end{eqnarray}
While calculating the polarization per site for spin model (\ref{207}), (\ref{209}) 
we have to use the energy spectrum (\ref{302}), (\ref{209})
and in addition to the contribution $p^y$ given in Eq.~(\ref{304}) 
to calculate the second term ${\sf{p}}^y$,
\begin{eqnarray}
\label{305}
{\sf{p}}={\sf{p}}^y\propto \frac{1}{N}\sum_n
\left(
2E\langle s^x_{n}s^z_{n+1}s^x_{n+2}+ s^y_{n}s^z_{n+1}s^y_{n+2}\rangle
\right.
\nonumber\\
\left.
-2K\langle s^x_{n}s^z_{n+1}s^y_{n+2}-s^y_{n}s^z_{n+1}s^x_{n+2}\rangle
\right)
\nonumber\\
=-\frac{1}{2\pi}\int_{-\pi}^{\pi}{\rm{d}}\kappa
\left[E\cos(2\kappa)-K\sin(2\kappa)\right] n_\kappa.
\end{eqnarray}

Furthermore,
we are interested in the entropy (per site) for the spin models at hand. 
It immediately follows from Eq.~({\ref{301}) through the relation $s=-\partial f/\partial T$,
\begin{eqnarray}
\label{306}
s=
\frac{1}{2\pi}\int_{-\pi}^{\pi}{\rm{d}}\kappa
\left[
\ln\left(2\cosh\frac{\epsilon_\kappa}{2T}\right)
-
\frac{\epsilon_\kappa}{2T}\tanh\frac{\epsilon_\kappa}{2T}
\right].
\end{eqnarray}

As emphasized in the end of the previous section,
the introduced three-spin interactions are essential for appearance of the nontrivial magnetoelectric effect.
This is clearly seen from formulas (\ref{302}) -- (\ref{304}) in the fermionic picture.
Obviously, for $T=0$ we have $n_\kappa=0$ if $\epsilon_\kappa>0$ and $n_\kappa=1$ if $\epsilon_\kappa<0$.
Therefore, for $E=K=0$
we have $\int_{-\pi}^{\pi}{\rm{d}}\kappa n_\kappa=\pi$ if ${\cal{H}}=0$ independently on ${\cal{E}}$,
and hence the ground-state magnetization (\ref{303}) is zero.
Furthermore, for $E=0$ 
we have $n_\kappa=n_{-\kappa}$ if ${\cal{E}}=0$ independently on ${\cal{H}}$,
and hence the ground-state polarization (\ref{304}) is zero.
In the presence of the three-site interactions the used symmetries of $\epsilon_\kappa$ (\ref{302}) may be broken
and the obtained conclusions do not hold any more.

For the sake of simplicity in the following analysis we distinguish two different types of three-spin interactions,
$XZY-YZX$ type, i.e., $K=0$ (Sec.~\ref{sec4})
and
$XZX+YZY$ type, i.e., $E=0$ (Sec.~\ref{sec5}).
From previous studies,\cite{rossler,drd,titvinidze,lou,kdsv08,topilko}
we know that for ${\cal{E}}=0$ both models 
[see Eqs.~(\ref{201}) -- (\ref{204})]
exhibit three phases in the ground state:
High-field ferromagnetic phase
and, depending on the relation between two- and three-spin interactions,
two different spin-liquid phases.
In the fermionic picture different phases correspond to different Fermi-surface topology 
(different number of Fermi points)
for fermions.
Furthermore, these models are known to exhibit a nonzero magnetization in zero magnetic field.\cite{rossler,drd}
Clearly, for ${\cal{E}}\ne 0$ 
both types of the three-spin interactions are present in the analysis of the magnetoelectric effect.

\section{Three-spin interactions of $XZY-YZX$ type}
\label{sec4}

In this section we consider the case of the $XZY-YZX$ three-spin interaction,
i.e., $E\ne 0$, $K=0$ ($J=1$).
From the ground-state phase diagram of the model (\ref{207}) with ${\cal{E}}=0$
(see, e.g., Ref.~\onlinecite{topilko})
we know that it makes sense to distinguish three representative values of $E$,
for example,
$E=0.5$, 
$E=1$,
and 
$E=2$.
For these cases we calculate 
the ground-state phase diagram in the plane ${\cal{H}}$--${\cal{E}}$ 
which indicates phases having different numbers of Fermi points
[see Fig.~\ref{fig01} for model (\ref{207}) with $E=2$],
and the ground-state\cite{footnote} magnetization and polarization
[see Fig.~\ref{fig02} for model (\ref{207}) with $E=0.5$ and $E=2$: 
$m({\cal{H}})$ (bold solid), $p({\cal{H}})$ (bold dashed) at ${\cal{E}}=0$ 
and
$m({\cal{E}})$ (bold solid), $p({\cal{E}})$ (bold dashed) at ${\cal{H}}=0$].
Furthermore, we calculate the low-temperature entropy in the plane ${\cal{H}}$--${\cal{E}}$ 
[see Fig.~\ref{fig03} for model (\ref{207}) with $E=2$ at $T=0.09$].

\begin{figure}
\begin{center}
\includegraphics[clip=on,width=6.5cm,angle=0]{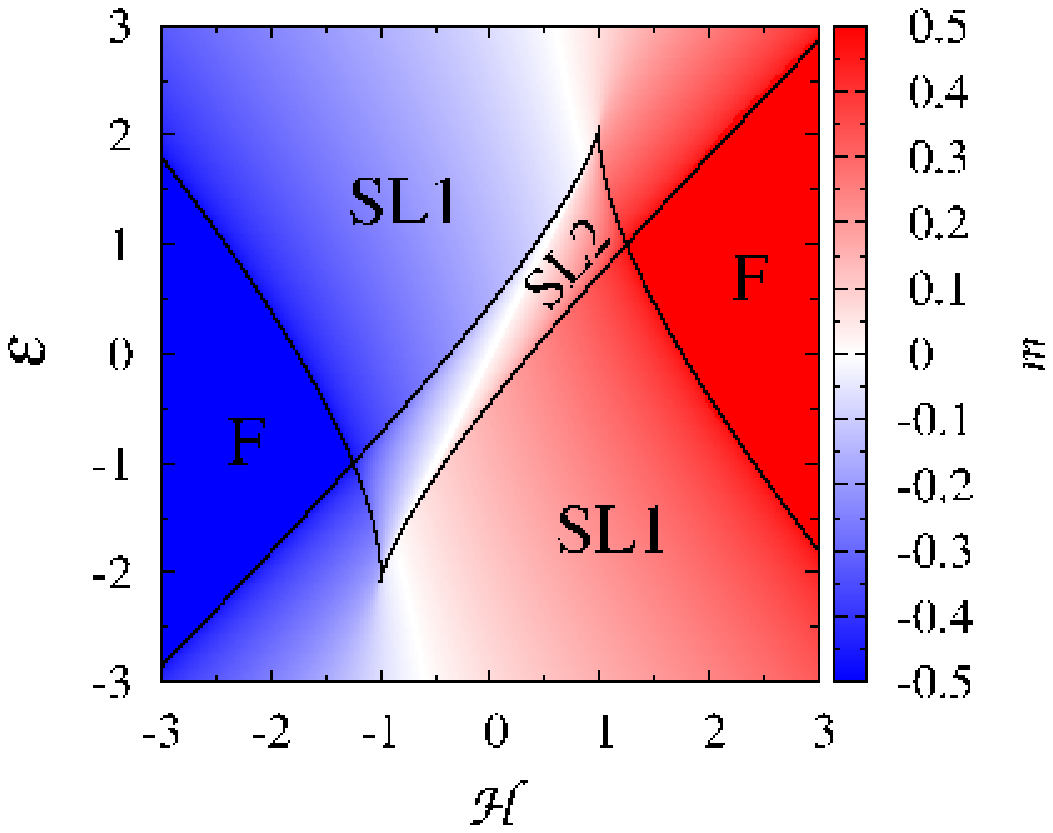}\\
\vspace{5mm}
\includegraphics[clip=on,width=6.5cm,angle=0]{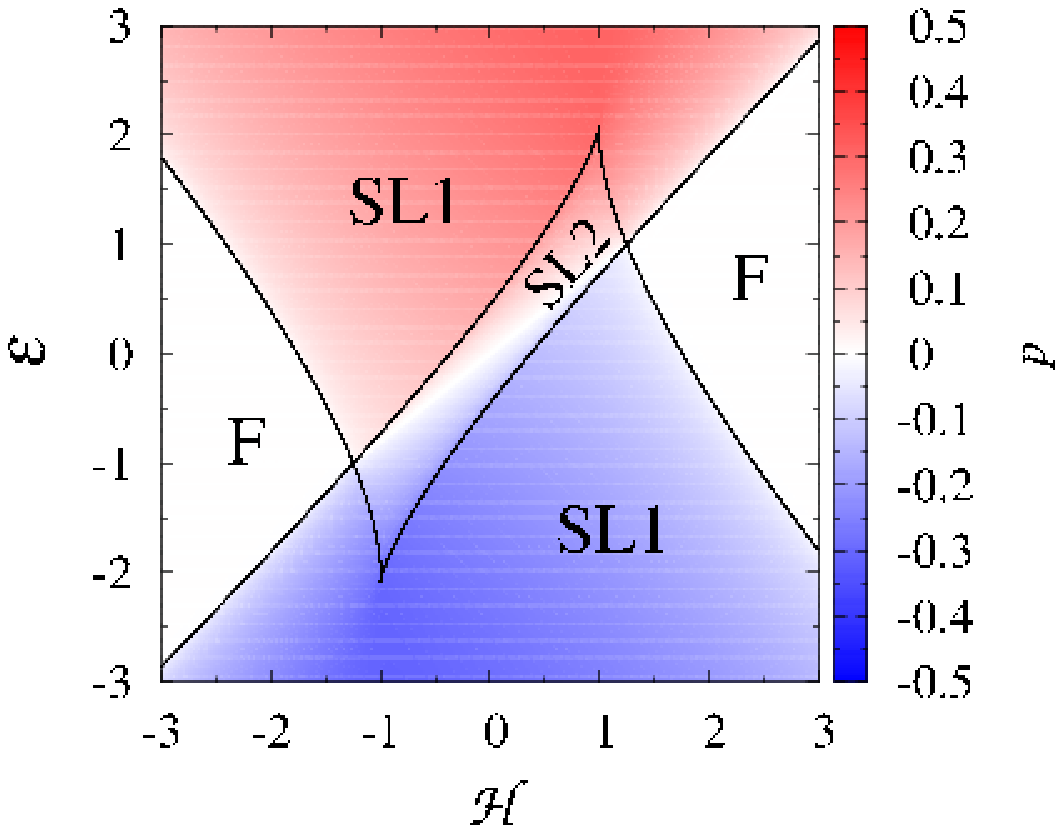}
\caption
{(Color online)
Magnetization (upper panel) 
and 
polarization (lower panel)
at very low temperature ($T=0.005$)
along with the ground-state phase diagram 
(black solid lines separate different phases)
of model (\ref{207}) with $J=1$, $E=2$, $K=0$.}
\label{fig01}
\end{center}
\end{figure} 

\begin{figure}
\begin{center}
\includegraphics[clip=on,width=4.25cm,angle=0]{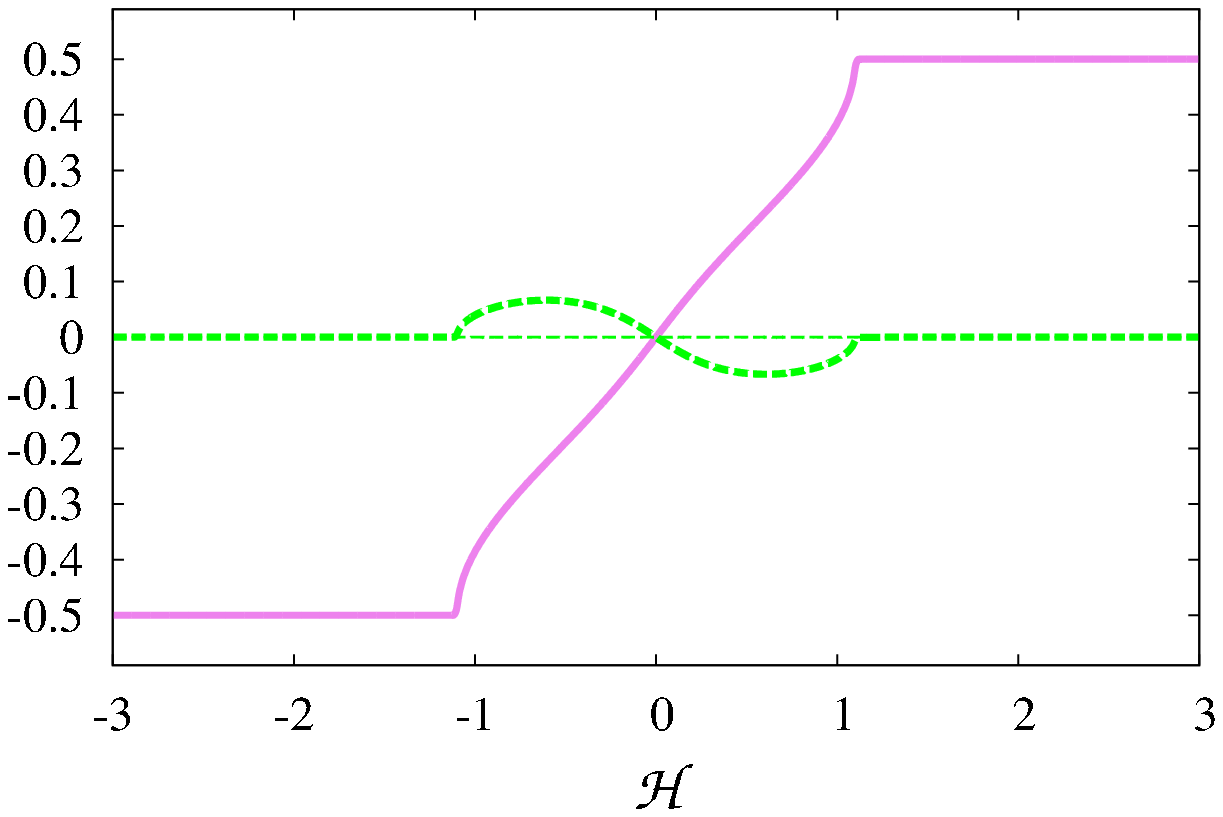}
\includegraphics[clip=on,width=4.25cm,angle=0]{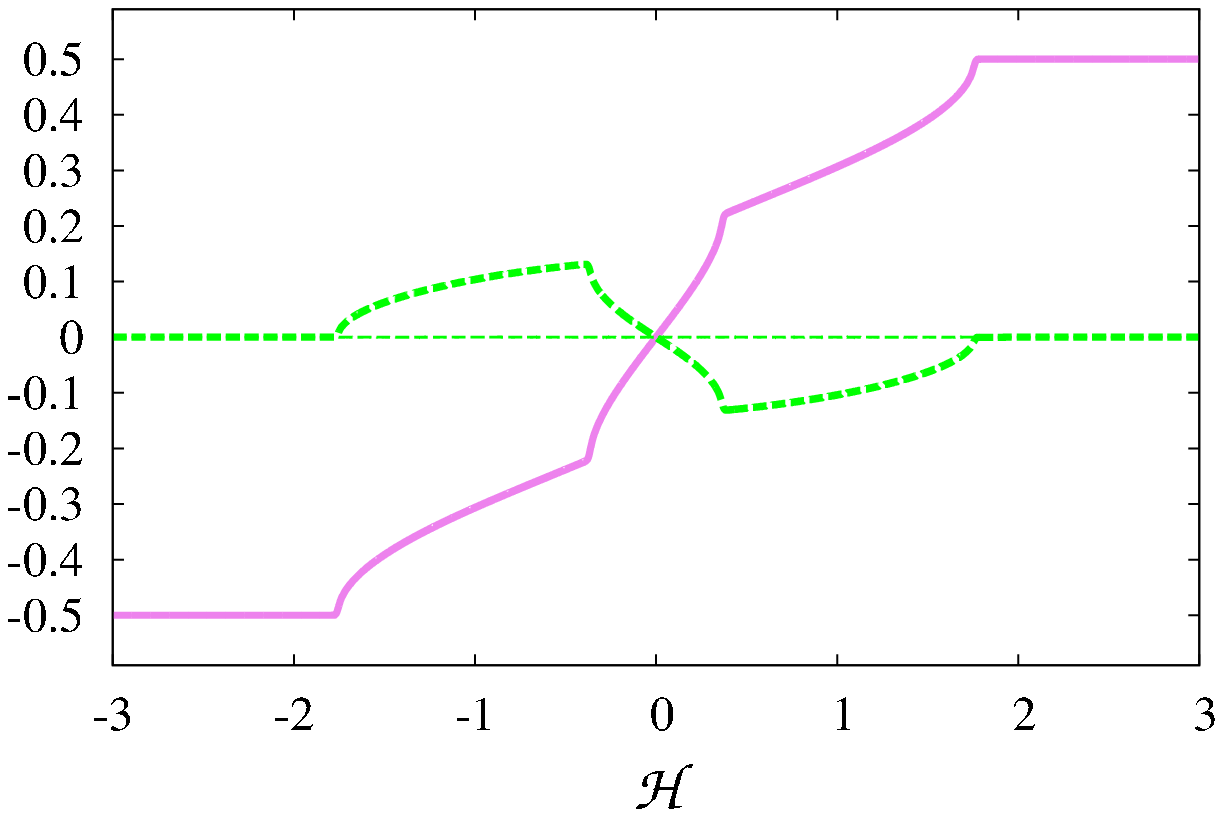}\\
\vspace{5mm}
\includegraphics[clip=on,width=4.25cm,angle=0]{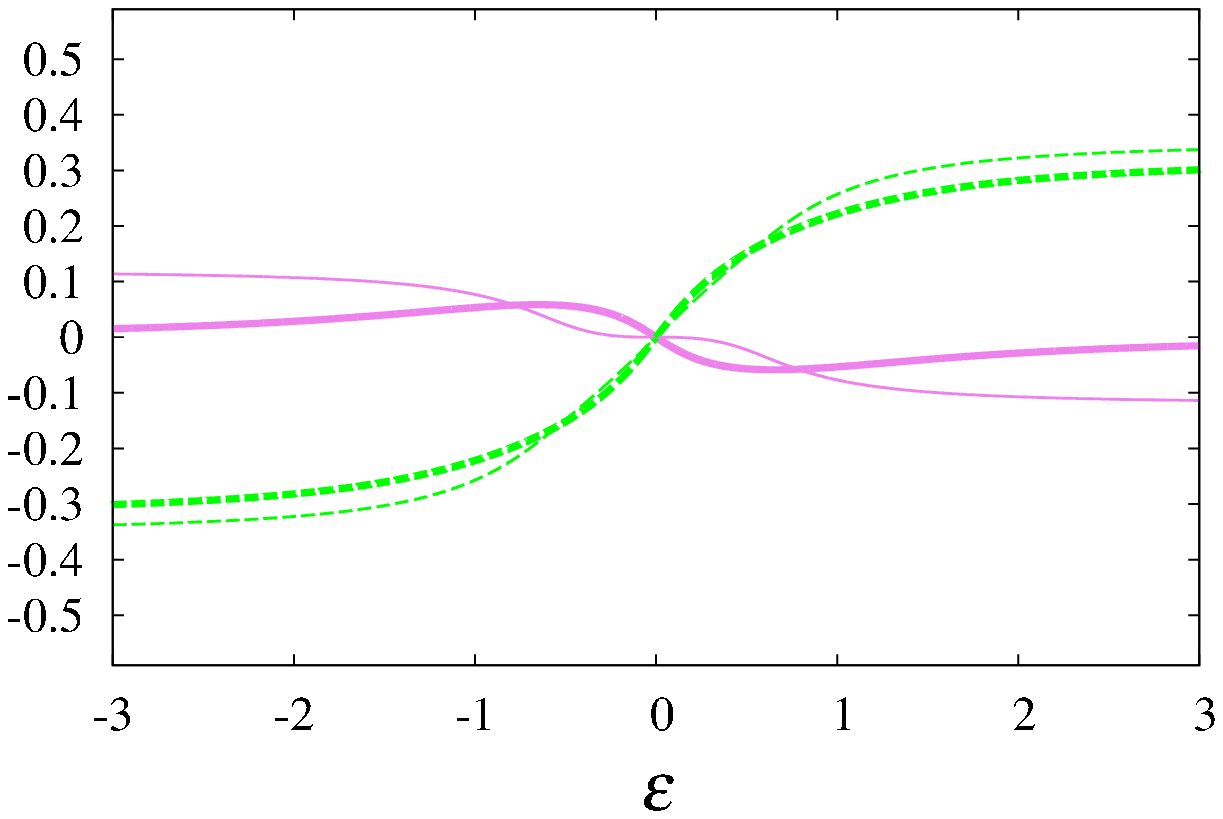}
\includegraphics[clip=on,width=4.25cm,angle=0]{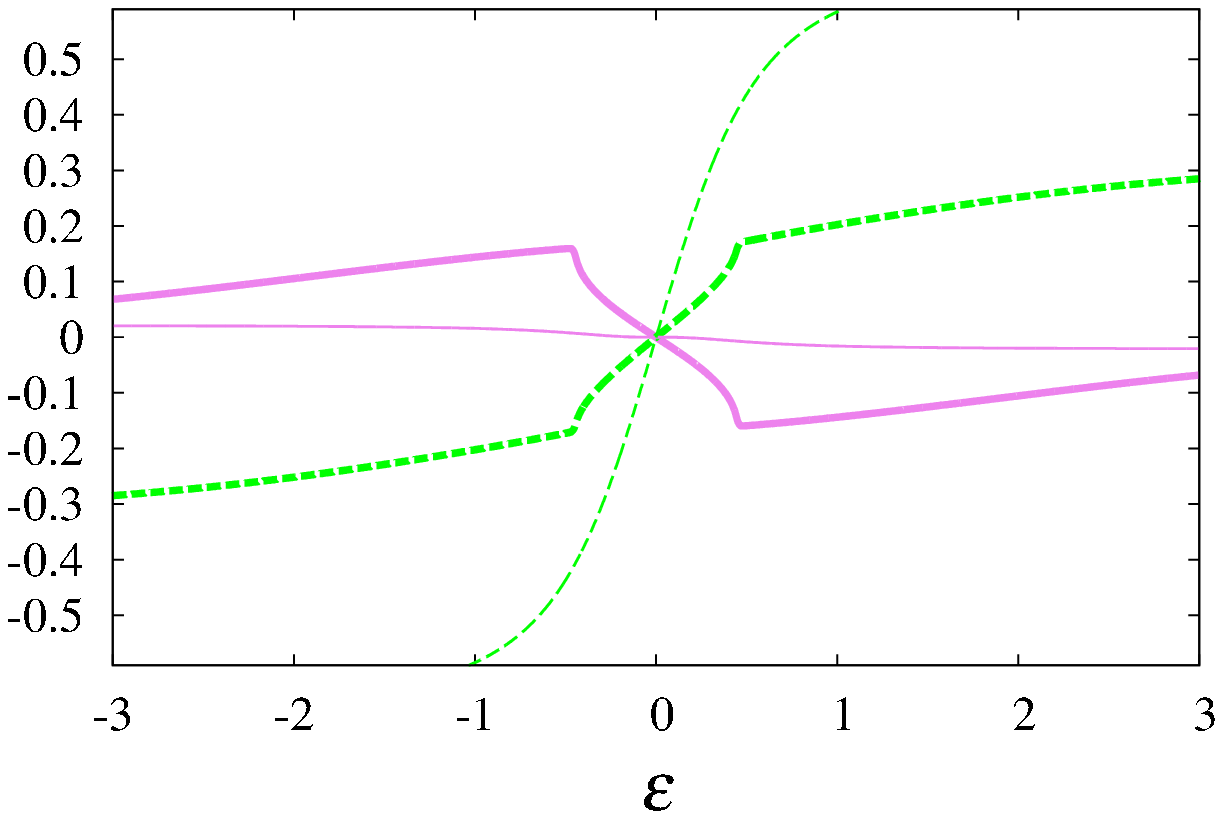}
\caption
{(Color online)
Dependencies of magnetization (bold solid) and polarization (bold dashed) at $T=0.005$
on magnetic field at ${\cal{E}}=0$ (first row)
and
on electric field at ${\cal{H}}=0$ (second row)
for model (\ref{207}) with $J=1$, $E=0.5$ (left column) or $E=2$ (right column), $K=0$.
By thin lines we show magnetization (thin solid) and polarization (thin dashed)
for model (\ref{207}), (\ref{209}) with $J=1$, $E=0.5,\,2$, $K=0$.}
\label{fig02}
\end{center}
\end{figure} 

\begin{figure}
\begin{center}
\includegraphics[clip=on,width=6.5cm,angle=0]{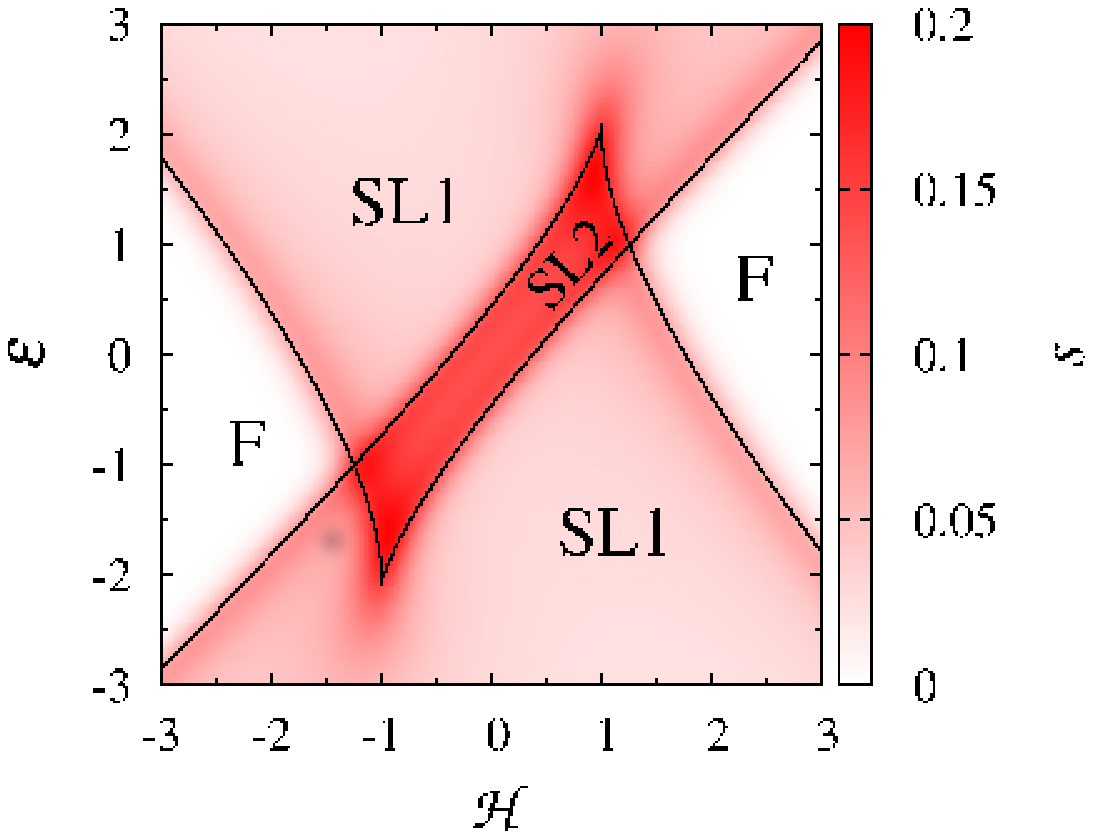}
\caption
{(Color online)
Towards magnetoelectrocaloric effect for model (\ref{207}) with $J=1$, $E=2$, $K=0$:
Entropy dependence on magnetic and electric fields at low temperature $T=0.09$
along with the ground-state phase diagram (black solid lines separate different phases).}
\label{fig03}
\end{center}
\end{figure} 

Let us discuss the obtained results.
The physics of the model along the line ${\cal{E}}=0$ in the plane ${\cal{H}}$--${\cal{E}}$ is well understood.\cite{rossler,lou,topilko}
While ${\cal{H}}$ increases from $-\infty$ to $\infty$
the system passes from the ferromagnetic phase to the ferromagnetic phase 
over the spin-liquid I phase if $\vert E\vert <1$
or
over the spin-liquid I phase, the spin-liquid II phase, and again the spin-liquid I phase if $\vert E\vert >1$.
The sequences of phases for the latter case with $E=2$ can be followed in Fig.~\ref{fig01} along the line ${\cal{ E}}=0$.
In the presence of the electric field ${\cal{E}}\ne 0$,
the system exhibits the same phases separated by quantum phase transition lines and two quantum triple point.
Ground-state dependences $m({\cal{H}})$ (solid lines) and $p({\cal{H}})$ (dashed lines) 
at ${\cal{E}}=0$ for $E=0.5$ and $E=2$ are shown in the first row in Fig.~\ref{fig02}.
These dependences are trivial only in the ferromagnetic phase,
when $m({\cal{H}})=\pm1/2$ and $p({\cal{H}})=0$.
In the both spin-liquid phases, 
not only the magnetization is influenced by the magnetic field 
but also the polarization even for ${\cal{E}}=0$ is effected by the magnetic field.
Furthermore,
from the ground-state dependences $m({\cal{E}})$ and $p({\cal{E}})$ at ${\cal{H}}=0$ 
(the second row in Fig.~\ref{fig02})
one can see that in the both spin-liquid phases not only $p$ depends on ${\cal{E}}$
but also $m$ even for ${\cal{H}}=0$ is governed by ${\cal{E}}$.

For the model at hand we can easily obtain rigorously further details about the behavior of relevant quantities.
Consider, e.g., the ground-state polarization $p({\cal{H}})$ 
slightly below the saturation field ${\cal{H}}_{\rm{sat}}>0$
for model (\ref{207}) with $J=1$, $E=0.5$ at ${\cal{E}}=0$,
see the corresponding panel in Fig.~\ref{fig02}.
Clearly,
the saturation field 
${\cal{H}}_{\rm{sat}}$
is defined by the fermion energy spectrum (\ref{302}),
namely,
$-{\cal{H}}_{\rm{sat}}+\cos\kappa^*-0.25\sin(2\kappa^*)=0$,
where $\kappa^*$ is determined from the equation
$\partial\epsilon_\kappa/\partial\kappa\vert_{\kappa=\kappa^*}=-\sin\kappa^*-0.5\cos(2\kappa^*)=0$,
i.e.,
$\kappa^*\approx -0.374\,734$.
Therefore ${\cal{H}}_{\rm{sat}}\approx 1.100\,917$.
Assume further ${\cal{H}}={\cal{H}}_{\rm{sat}}-\delta {\cal{H}}$, where $\delta {\cal{H}}>0$ is a small quantity.
According to Eq.~(\ref{304}),
$p=(\cos\kappa_1^*-\cos\kappa_2^*)/(2\pi)$,
where 
$\kappa^*_1\approx \kappa^*-1.113\,915\sqrt{\delta {\cal{H}}}$
and
$\kappa^*_2\approx \kappa^*+1.113\,915\sqrt{\delta {\cal{H}}}$.
As a result, 
at ${\cal{H}}={\cal{H}}_{\rm{sat}}-\delta {\cal{H}}$ we have
$p({\cal{H}})
\approx
-0.129\,782\sqrt{{\cal{H}}_{\rm{sat}}-{\cal{H}}}$,
that is, the ground-state polarization emerges with the critical exponent 1/2 
as the control parameter ${\cal{H}}$ passes the critical value ${\cal{H}}_{\rm{sat}}$.

The ground-state dependences of $m$ and $p+{\sf{p}}$ on ${\cal{H}}$ and ${\cal{E}}$ for model (\ref{207}), (\ref{209})
are shown in Fig.~\ref{fig02} by thin lines.
Within the fermionic picture (\ref{302}), (\ref{304}), (\ref{305}) it can be proved
(Appendix~\ref{b})
that for this model
$p+{\sf{p}}=0$ if ${\cal{E}}=0$ independently on ${\cal{H}}$,
see thin dashed lines in the two panels from the first row in Fig.~\ref{fig02}.
For nonzero ${\cal{E}}$,
however,
$p+{\sf{p}}$ is influenced by ${\cal{H}}$.
Clearly, $m$ at ${\cal{E}}=0$ 
for the model (\ref{207}) and the model (\ref{207}), (\ref{209}) is the same.

The ground-state phase diagram also manifests itself 
in the dependence of the entropy $s$ (\ref{306}) on ${\cal{H}}$ and ${\cal{E}}$ at low temperatures.
The low-temperature entropy exhibits well pronounced maxima 
along the quantum phase transition lines and around the quantum triple points.
These maxima become more sharper as the temperature decreases.
If the system is placed in a thermostat with the temperature $T$,
$\Delta Q=T\,\Delta S$ with $\Delta S=S({\cal{H}}_2,{\cal{E}}_2)-S({\cal{H}}_1,{\cal{E}}_1)$
is the heat the system takes in (if $\Delta S>0$) or gives out (if $\Delta S<0$)
under the change of the fields from the values ${\cal{H}}_1,{\cal{E}}_1$ to the values ${\cal{H}}_2,{\cal{E}}_2$.
Clearly,
the system at hand exhibits a magnetoelectrocaloric effect,
i.e.,
can be used for cooling/heating under a change of external fields.
The magnetoelectrocaloric effect is most pronounced 
at low temperatures 
around the quantum phase transition lines and around the quantum triple points,
see Fig.~\ref{fig03}.

\section{Three-spin interactions of $XZX+YZY$ type}
\label{sec5}

We pass to the case of the $XZX+YZY$ three-spin interaction,
i.e., $E=0$, $K\ne 0$ ($J=1$).
From the ground-state phase diagram of the model (\ref{207}) with ${\cal{E}}=0$
(see, e.g., Ref.~\onlinecite{topilko})
we know that it makes sense to distinguish three representative values of $K$,
for example,
$K=0.25$,
$K=0.5$,
and 
$K=1.7$.
For these cases we calculate 
the ground-state phase diagram in the plane ${\cal{H}}$--${\cal{E}}$ 
which indicates phases having different numbers of Fermi points
[see Fig.~\ref{fig04} for model (\ref{207}) with $K=1.7$],
and the ground-state\cite{footnote} magnetization and polarization
[see Fig.~\ref{fig05} for model (\ref{207}) with $K=0.5$ and $K=1.7$:
$m({\cal{H}})$ (bold solid), $p({\cal{H}})$ (bold dashed) at ${\cal{E}}=0$ 
and
$m({\cal{E}})$ (bold solid), $p({\cal{E}})$ (bold dashed) at ${\cal{H}}=0$].
Furthermore, we calculate the low-temperature entropy in the plane ${\cal{H}}$--${\cal{E}}$
[see Fig.~\ref{fig06} for model (\ref{207}) with $K=1.7$ at $T=0.09$].

\begin{figure}
\begin{center}
\includegraphics[clip=on,width=6.5cm,angle=0]{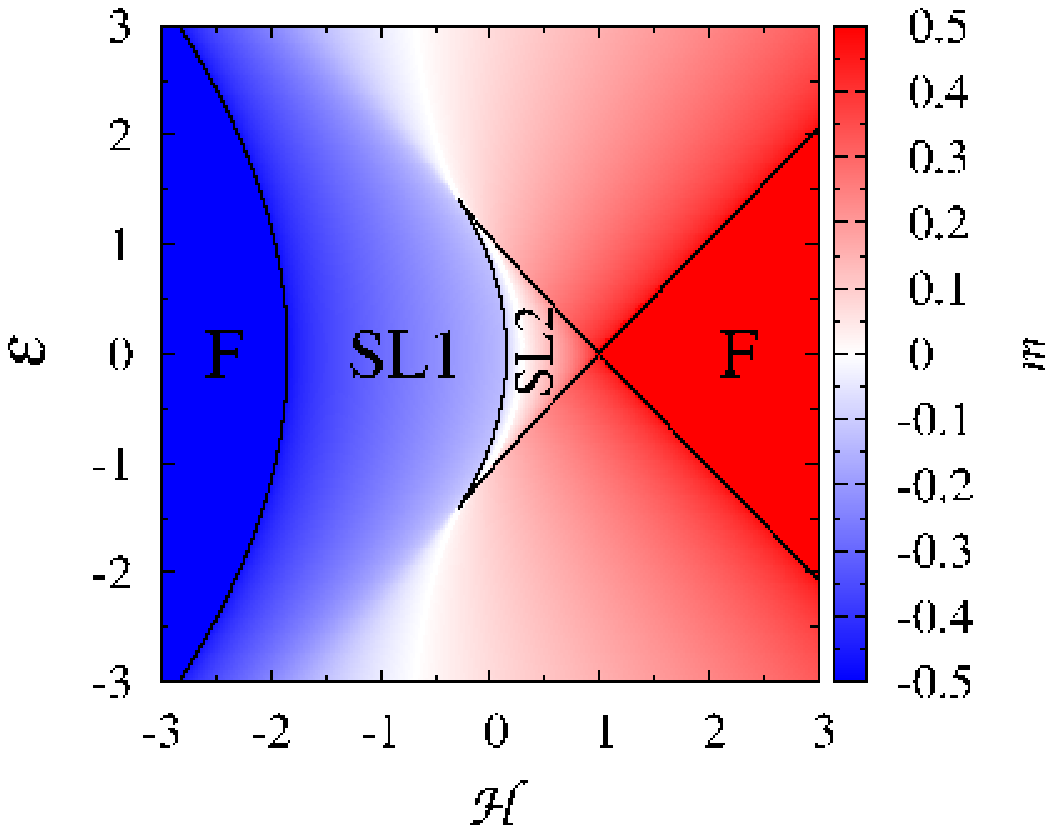}\\
\vspace{5mm}
\includegraphics[clip=on,width=6.5cm,angle=0]{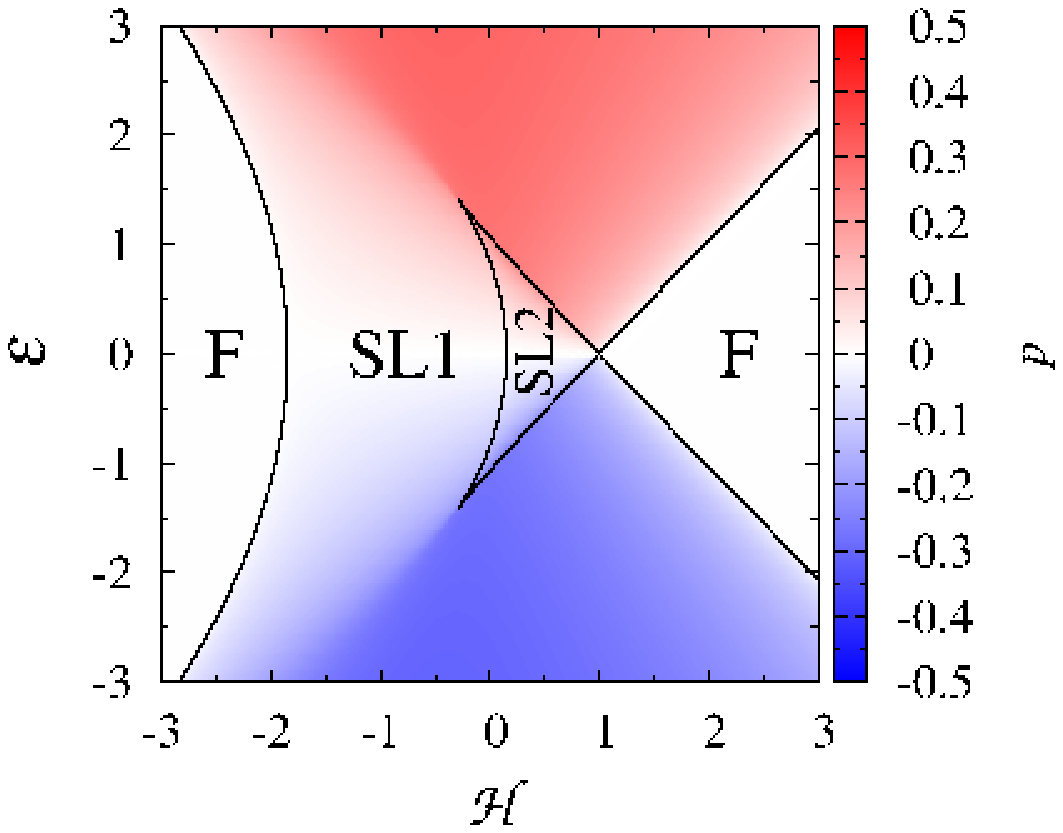}
\caption
{(Color online)
Magnetization (upper panel) 
and 
polarization (lower panel)
at very low temperature ($T=0.005$)
along with the ground-state phase diagram 
(black solid lines separate different phases)
of model (\ref{207}) with $J=1$, $E=0$, $K=1.7$.}
\label{fig04}
\end{center}
\end{figure} 

\begin{figure}
\begin{center}
\includegraphics[clip=on,width=4.25cm,angle=0]{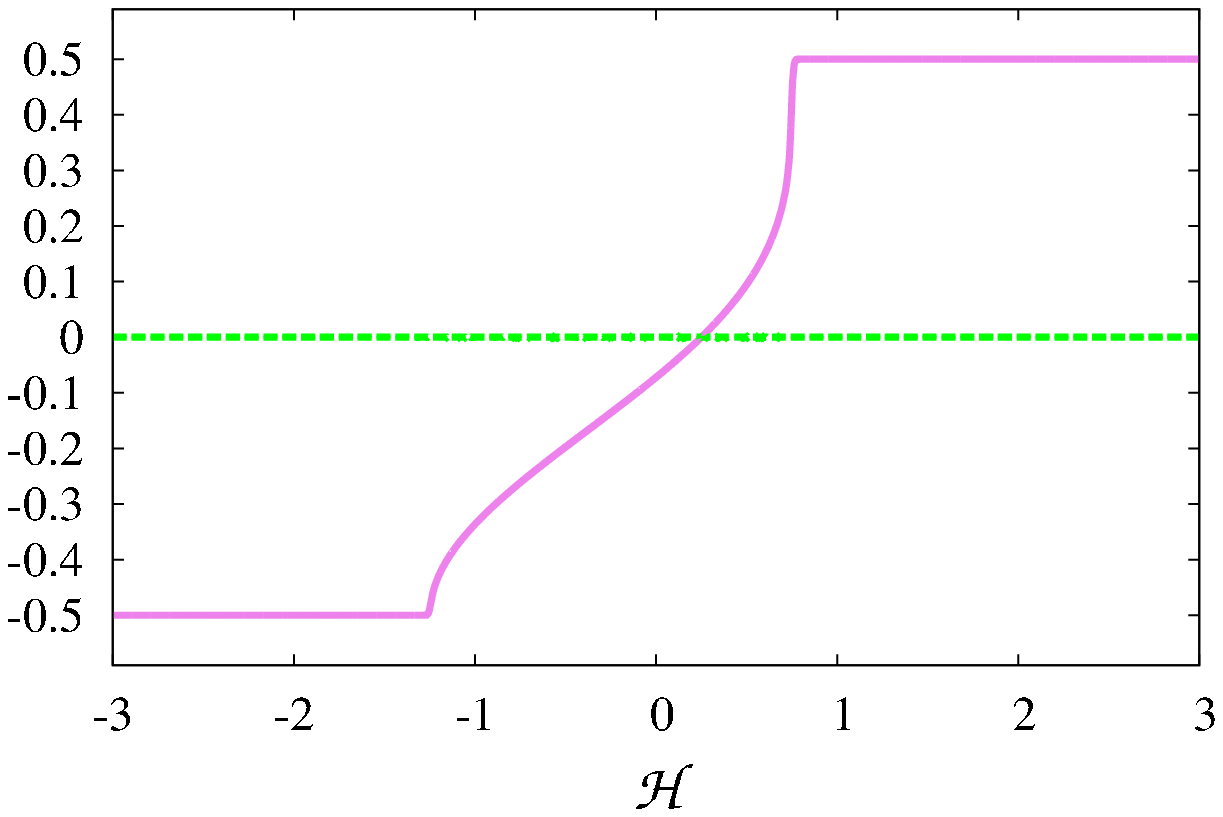}
\includegraphics[clip=on,width=4.25cm,angle=0]{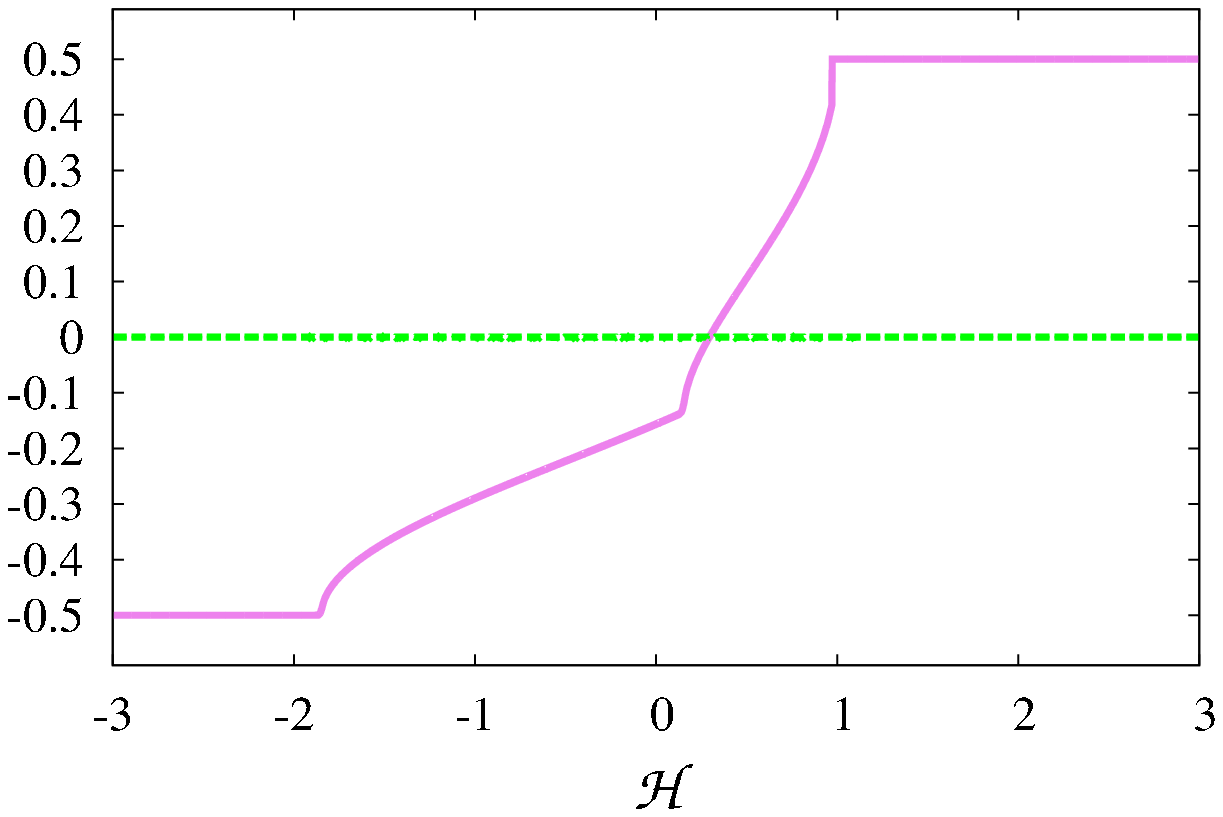}\\
\vspace{5mm}
\includegraphics[clip=on,width=4.25cm,angle=0]{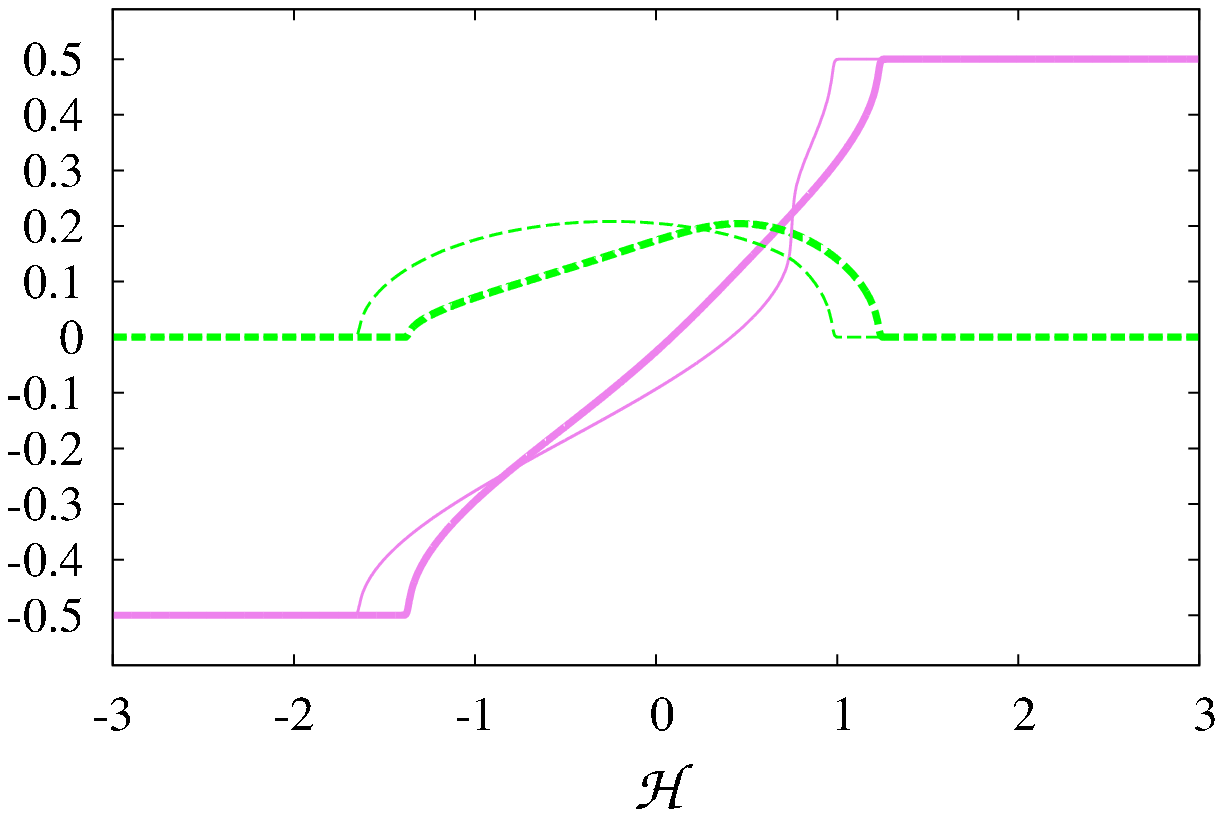}
\includegraphics[clip=on,width=4.25cm,angle=0]{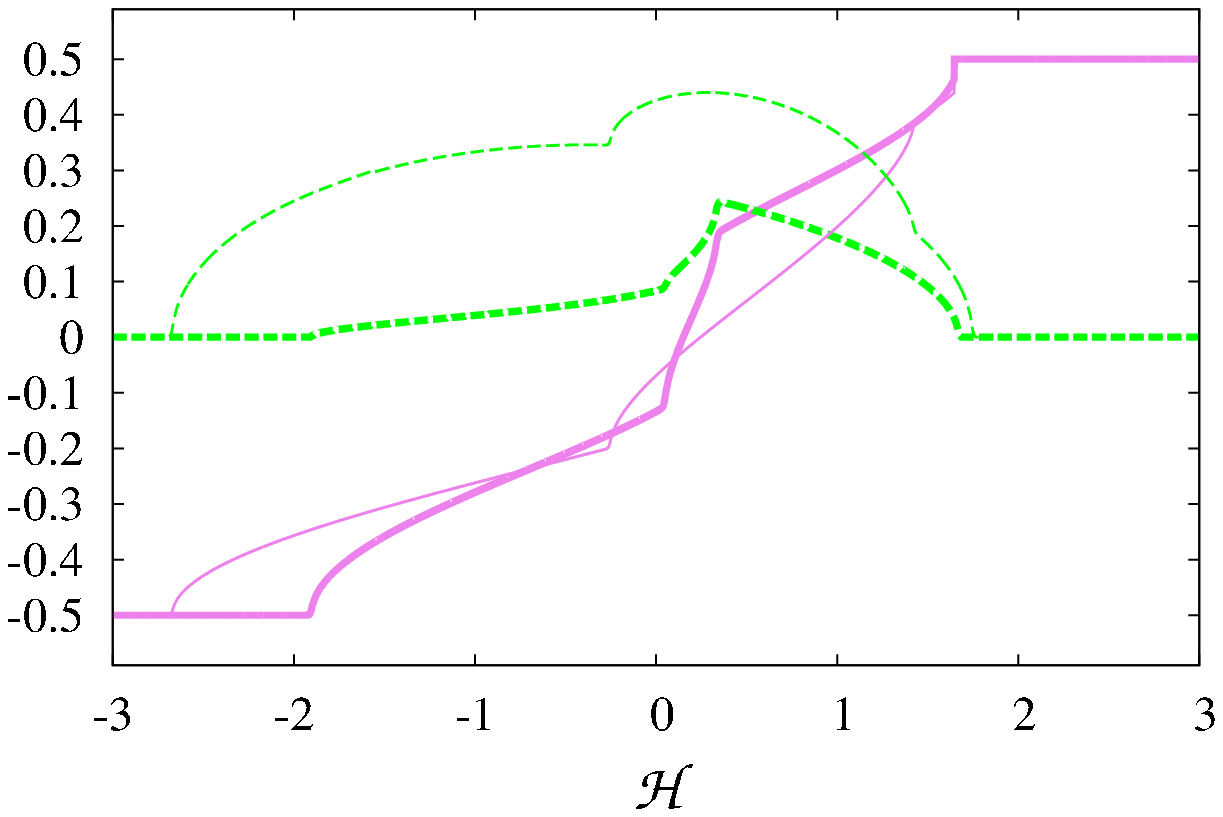}\\
\vspace{5mm}
\includegraphics[clip=on,width=4.25cm,angle=0]{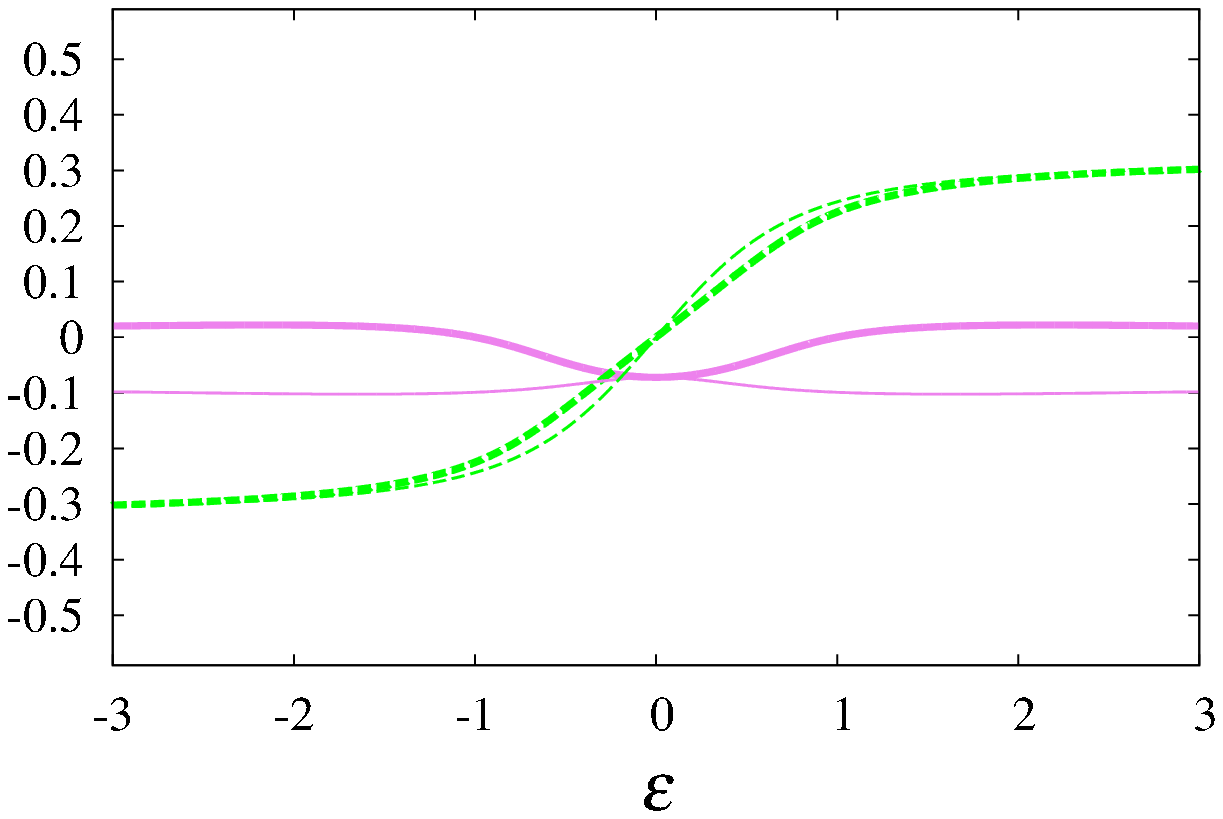}
\includegraphics[clip=on,width=4.25cm,angle=0]{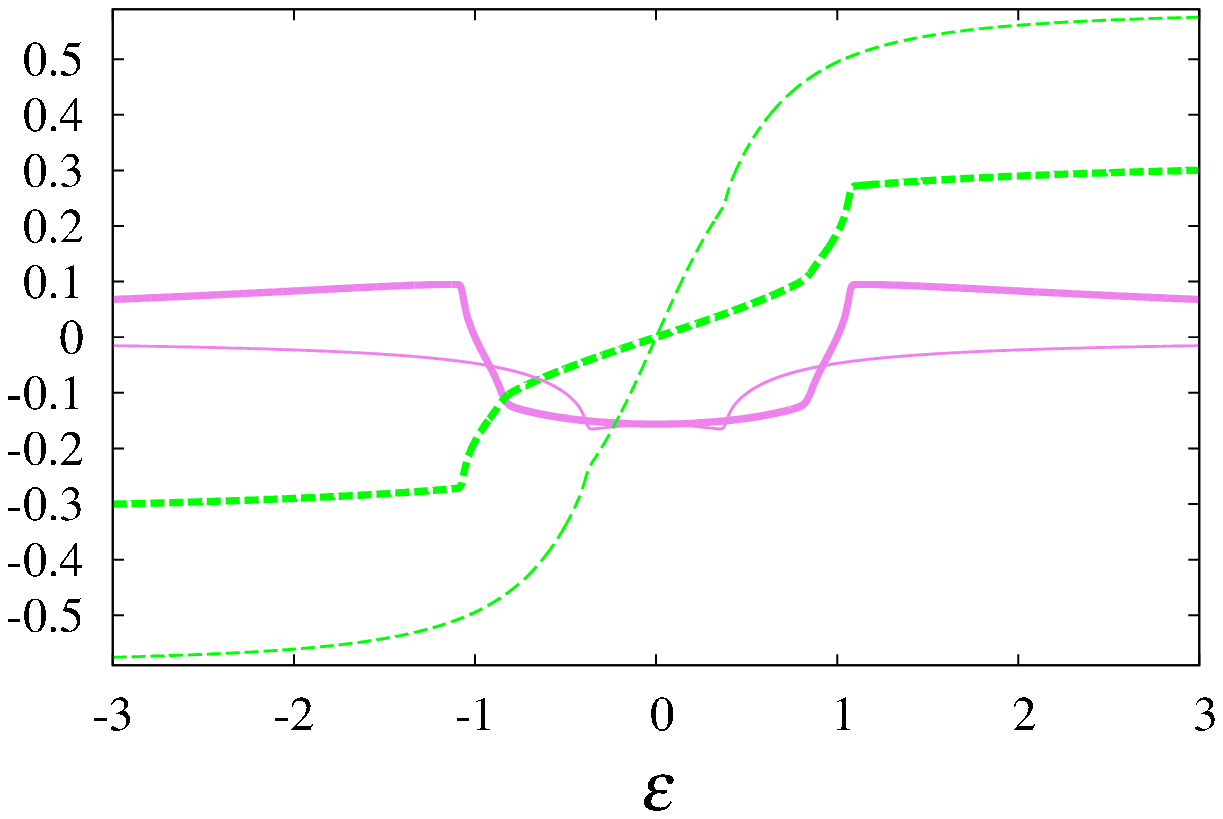}
\caption
{(Color online)
Dependencies of magnetization (bold solid) and polarization (bold dashed) at $T=0.005$
on magnetic field at ${\cal{E}}=0$ (first row) and ${\cal{E}}=0.7$ (second row)
and
on electric field at ${\cal{H}}=0$ (third row)
for model (\ref{207}) with $J=1$, $E=0$, $K=0.5$ (left column) or $K=1.7$ (right column).
By thin lines we show magnetization (thin solid) and polarization (thin dashed)
for model (\ref{207}), (\ref{209}) with $J=1$, $E=0$, $K=0.5,\,1.7$.}
\label{fig05}
\end{center}
\end{figure} 

\begin{figure}
\begin{center}
\includegraphics[clip=on,width=6.5cm,angle=0]{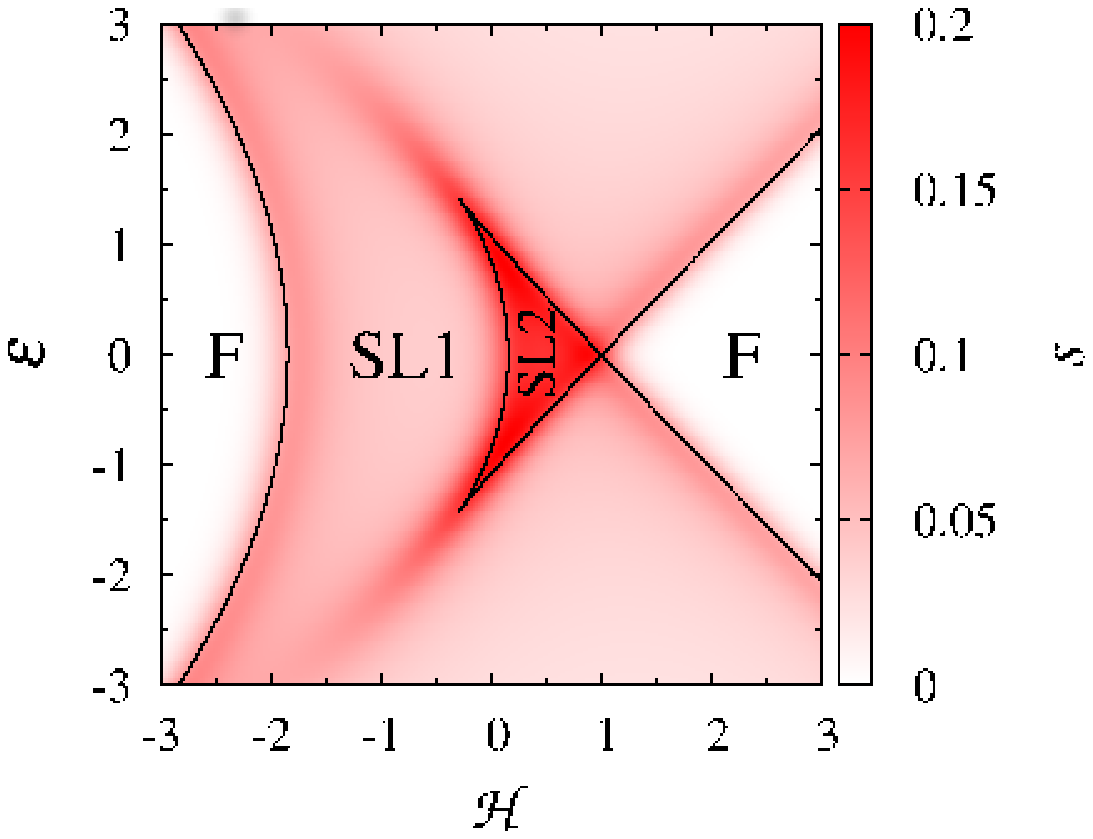}
\caption
{(Color online)
Towards magnetoelectrocaloric effect for model (\ref{207}) with $J=1$, $E=0$, $K=1.7$:
Entropy dependence on magnetic and electric fields at low temperature $T=0.09$
along with the ground-state phase diagram (black solid lines separate different phases).}
\label{fig06}
\end{center}
\end{figure} 

Again the properties of the model along the line ${\cal{E}}=0$ are well known.\cite{titvinidze,topilko}
As ${\cal{H}}$ varies from $-\infty$ to $\infty$,
the system is driven by varying ${\cal{H}}$ from the ferromagnetic phase to the ferromagnetic phase
through the spin-liquid I phase if $\vert K\vert<1/2$
or 
through the spin-liquid I and spin-liquid II phases 
(through the spin-liquid II and spin-liquid I phases)
if $K>1/2$
(if $K<-1/2$).
Ground-state dependences of $m$ and $p$ (and $p+{\sf{p}}$) on fields
show nontrivial features only outside the ferromagnetic phase.
However,
$p({\cal{H}})=0$ along the line ${\cal{E}}=0$ that is obviously traced back to the symmetry
$\epsilon_\kappa=\epsilon_{-\kappa}$,
see the first row in Fig.~\ref{fig05}.
Breaking this symmetry by switching on ${\cal{E}}$ immediately results in influence of ${\cal{H}}$ on $p$,
see the second row in Fig.~\ref{fig05}.
On the other hand,
$m$ is influenced by ${\cal{E}}$ even at ${\cal{H}}=0$,
see the third row in Fig.~\ref{fig05}.
Again the considered model exhibits a magnetoelectrocaloric effect,
which is most pronounced at low temperatures around characteristic lines of the ground-state phase diagram,
see Fig.~\ref{fig06}.

\section{Conclusions}
\label{sec6}

We have considered simple but nontrivial models of a multiferroic of spin origin.
The main worth of the models is their exact solvability:
All relevant quantities can be calculated rigorously and examined in detail.
These studies may serve as a benchmark for more realistic cases which are not exactly solvable.

In contrast to free-fermion models studied earlier,\cite{vadim,thakur,oles}
we include in the model the three-spin interactions of $XZY-YZX$ and $XZX+YZY$ types.
Due to these interactions the magnetoelectric effect becomes especially interesting:
Magnetization (polarization) can be induced and governed solely by electric (magnetic) field.
The considered models show magnetoelectrocaloric effect,
i.e.,
isothermally (adiabatically) varying fields noticeably change the entropy (temperature).
The effect is most pronounced at low temperatures around peculiarities 
(quantum phase transition lines and quantum triple points)
on the ground-state phase diagram
which is rather rich in the presence of the three-spin interactions.

It should be stressed 
that some characteristic features of the reported dependences for the polarization or the magnetization on fields
(e.g., emergence after passing a threshold value, cusps, abrupt changes, etc.)
can be seen in experimentally measured data,
see Ref.~\onlinecite{mee2} and references therein.
On the other hand,
the minimal model to describe such spin-chain multiferroics as LiCu$_2$O$_2$ or LiCuVO$_4$
is the spin-1/2 anisotropic Heisenberg model with the Hamiltonian
\begin{eqnarray}
\label{601}
H_0=\sum_{n}
\left[
J_1\left(s_n^xs_{n+1}^x+s_n^ys_{n+1}^y +\Delta s_n^zs_{n+1}^z\right)
\right.
\nonumber\\
\left.
+
J_2\left(s_n^xs_{n+2}^x+s_n^ys_{n+2}^y +\Delta s_n^zs_{n+2}^z\right)
\right.
\nonumber\\
\left.
-{\cal{H}}s_n^z
\right],
\end{eqnarray}
with ferromagnetic $J_1<0$, antiferromagnetic $J_2>0$, 
and small easy-plane anisotropy $\Delta\le 1$,
see, e.g., Ref.~\onlinecite{sirker}.
One way to examine this model is to apply the Jordan-Wigner fermionization method.
However, in contrast to the model given in Eqs.~(\ref{201}) -- (\ref{204}),
one arrives for model (\ref{601}) at interacting fermions and further treatment becomes approximate. 

Finally, 
the considered models hold promise as a core system permitting to examine some other aspects of multiferroics,
e.g., the dynamical magnetoelectric effect.\cite{tokura}
The work in this direction is in progress.

\section*{Acknowledgments}

The authors thank J.~Richter for discussions.
The present study was supported by the ICTP (OEA, network-68):
T.~V. acknowledges the kind hospitality of the Yerevan University
and
V.~O. acknowledges the kind hospitality of the ICMP in 2014;
O.~M. and T.~V. acknowledge the kind hospitality of the Yerevan University in 2015.
V.~O. also acknowledges the partial financial support form the grant 
by the State Committee of Science of Armenia No.~13-1F343.
V.~O. and O.~D. are supported by the ICTP
through the Junior Associate award and Senior Associate award,
respectively.

\appendix
\section{Jordan-Wigner transformation and thermodynamic functions}
\label{a}

Let us consider statistical mechanical calculations for the spin-1/2 chain model given in Eq.~(\ref{207}).
First we introduce the operators $s_j^{\pm}=s_j^x\pm{\rm{i}}s_j^{y}$.
Then we use the Jordan-Wigner transformation to spinless fermions,
\begin{eqnarray}
\label{a01}
c^\dagger_1=s_1^+,
\;
c^\dagger_j=(-2s^z_1)\ldots(-2s^z_{j-1})s_j^+, 
j=2,\ldots,N,
\nonumber\\
c_1=s_1^-,
\;
c_j=(-2s^z_1)\ldots(-2s^z_{j-1})s_j^-, 
j=2,\ldots,N,
\end{eqnarray}
to get for the Hamiltonian (\ref{207}) a bilinear Fermi-form:
\begin{eqnarray}
\label{a02}
H=\sum_n
\left[
\frac{J+{\rm{i}}J{\cal{E}}}{2}c_n^\dagger c_{n+1} 
-\frac{{\rm{i}}E+K}{4}c_n^\dagger c_{n+2} + {\rm{H.c.}}
\right.
\nonumber\\
\left.
-{\cal{H}}\left(c_n^\dagger c_n-\frac{1}{2}\right)
\right].
\end{eqnarray}
This Hamiltonian can be brought into the diagonal form after the Fourier transformation
\begin{eqnarray}
\label{a03}
c_\kappa^\dagger=\frac{1}{\sqrt{N}}\sum_n e^{-{\rm{i}}\kappa n} c_n^\dagger,
\;
c_\kappa=\frac{1}{\sqrt{N}}\sum_n e^{{\rm{i}}\kappa n} c_n,
\end{eqnarray}
$\kappa=2\pi m/N$, $m=-N/2,\ldots, N/2-1$ 
(we assume without loss of generality that $N$ is even),
which yields
\begin{eqnarray}
\label{a04}
H=\sum_\kappa \epsilon_\kappa \left(c_\kappa^\dagger c_\kappa-\frac{1}{2}\right)
\end{eqnarray}
with $\epsilon_\kappa$ given in Eq.~(\ref{302}).

The partition function of $N$ Fermi oscillators (\ref{a04}) can be easily calculated,
$Z(T,{\cal{H}},{\cal{E}},N)=\prod_\kappa 2\cosh[\epsilon_\kappa/(2T)]$.
It yields the Helmholtz free energy per site given in Eq.~(\ref{301}).
To get the magnetization (\ref{303}) and the polarization (\ref{304}) 
one may simply take the corresponding derivatives,
i.e.,
$m=-\partial f/\partial {\cal{H}}$
and
$p=-\partial f/\partial {\cal{E}}$.

\section{Ground-state polarization for model (\ref{207}), (\ref{209}) at ${\cal{E}}=0$}
\label{b}

Consider the ground-state polarization $p+{\sf{p}}$ for model (\ref{207}), (\ref{209}).
According to Eqs. (\ref{304}), (\ref{305}) we have to calculate
\begin{eqnarray}
\label{b01}
-\frac{1}{2\pi}
\int_{\Gamma}
{\rm{d}}\kappa
\left[J\sin\kappa +E\cos(2\kappa)-K\sin(2\kappa)\right].
\end{eqnarray}
Here $\Gamma$ is the domain within the interval $[-\pi,\pi]$ 
where
$\epsilon_\kappa=-{\cal{H}}+J\cos\kappa-(E/2)\sin(2\kappa)-(K/2)\cos(2\kappa)<0$
(we consider the case ${\cal{E}}=0$).
The antiderivate in Eq.~(\ref{b01}) is
$-J\cos\kappa+(E/2)\sin(2\kappa)+(K/2)\cos(2\kappa)=-\epsilon_\kappa-{\cal{H}}$
and therefore the resulting integral is zero
since $\epsilon_\kappa=0$ at the boundaries of the domain $\Gamma$.
Moreover,
$p+{\sf{p}}=0$ for model (\ref{207}), (\ref{209}) at ${\cal{E}}=0$
for nonzero temperatures too.


\begin{thebibliography}{999}
\bibitem{footnote}
Although the results reported in Figs.~\ref{fig02} and \ref{fig05} refer to a very small but nonzero temperature ($T=0.005$)
they can not be distinguished from the results at $T=0$ in the scale of the figures.
\end{thebibliography}
\end{document}